\def\f#1#2{{\textstyle{#1\over#2}}}        
\def\({\eqno(}
\def\on#1#2{{\buildrel{\mkern2.5mu#1\mkern-2.5mu}\over{#2}}}
\def\Dot#1{\on{\hbox{\bf .}}{#1}}       \def\dot{\Dot}
\def\sect#1\par{\par\ifdim\lastskip<\medskipamount
        \bigskip\medskip\goodbreak\else\nobreak\fi
        \noindent{\large\bf{#1}}\par\nobreak\medskip}
\def\refs{\sect{REFERENCES}\par\medskip \frenchspacing 
        \parskip=0pt \small
\renewcommand{\baselinestretch}{1}}
\def\Item#1 {\par\hang\textindent{[#1]}}
\parskip=\medskipamount 
\newskip\humongous \humongous=0pt plus 1000pt minus
1000pt

\newif\ifdtup

\documentstyle[12pt]{article}

\def\de{\nabla}  

\def\Bar#1{\overline{#1}}                       

\renewcommand{\a}{\alpha}
\renewcommand{\b}{\beta}

\renewcommand{\d}{\delta}
\newcommand{\q}{\theta}

\newcommand{\g}{\gamma}

\newcommand{\e}{\epsilon}

\newcommand{\m}{\mu}

\newcommand{\ad}{{\Dot{\alpha}}}
\newcommand{\bd}{{\Dot{\beta}}}
\newcommand{\gd}{{\Dot{\gamma}}}
\newcommand{\Del}{\nabla}

\begin{document}

\begin{titlepage}
\begin{flushright} {\hbox to\hsize{November 1997 \hfill
hep-th/9711151}} {UMDEPP 98-47}\\ {BRX-TH-421}\\
{McGill/97-31}\\ {ITP-SB-97-65}
\end{flushright}
\vspace{.2cm}
\begin{center}

{\Large\bf  Component Actions from Curved
Superspace:\\[0.05in] Normal Coordinates and
Ectoplasm\footnote{ Research supported by NSF grants
PHY-96-43219, PHY-9604587, and PHY-9722101.}}
\\[.24in] S. James Gates,
Jr.\footnote{gates@umdhep.umd.edu} \\[.04in] {\it
Department of Physics, University of Maryland at College
Park\\ College Park, MD 20742-4111 USA} 
\\[.06in] Marcus T.
Grisaru\footnote{grisaru@binah.cc.brandeis.edu}\\[.04in]
{\it  Physics Department, Brandeis University\\ Waltham,
MA\quad 02254 USA}\\ [.06in] Marcia E.
Knutt-Wehlau\footnote{knutt@physics.mcgill.ca.  Supported
in part by a  NSERC Postdoctoral Fellowship.}\\[.04in] {\it
Physics Department, McGill University\\ Montreal, PQ
CANADA H3A 2T8}\\ {\rm {and}}\\[.06in] Warren 
Siegel\footnote{siegel@insti.physics.sunysb.edu}\\[.04in]
{\it Institute for Theoretical Physics, State University of
New York\\ Stony Brook, NY 11794-3840 USA}

\vspace{.25in} {ABSTRACT}\\[.1in]
\end{center}
\begin{quote} We give efficient superspace methods for
deriving component actions  for supergravity coupled to
matter.  One method uses normal coordinates to 
covariantly expand the superfield action, and can be
applied  straightforwardly to any superspace.  The other
interprets the  component lagrangian as a differential form
on a bosonic hypersurface  in superspace, and gives a simple
derivation for pertinent  cases such as chiral superspace.
\end{quote}

\end{titlepage}

\sect{1. Introduction}

Superspace methods have many advantages over component
approaches  to supersymmetry (see, e.g., [1]).  They allow
us to write manifestly supersymmetric actions, and they
facilitate quantum calculations. However,  component
expansions of superfields  and superspace actions can be
useful  for  comparison to nonsupersymmetric theories,  or
for applying supersymmetry to nonsupersymmetric theories.

Although component expansions of superspace actions are
easy and straightforward for globally supersymmetric
theories, the same has not been true for locally
supersymmetric ones.  The two main complications are: (1)
the complexity of supergravity in comparison to super
Yang-Mills theory, as reflected in the commutation relations
of the covariant derivatives needed for  component
expansions, and (2) the lack of a  simple way to expand the
integration measure, which is absent in the global case.
Consequently, the construction of locally supersymmetric
component actions, either by starting from superspace, or 
directly by requiring component local supersymmetry, has
always been a somewhat awkward marriage of various ad
hoc techniques.

This letter gives  a brief overview of two recent papers
[2,3] that have  tackled the problem anew, using two
complementary approaches.   One approach is based on
component expansions  with respect to fermionic Riemann 
normal coordinates [4,5], equivalent to covariant
Wess-Zumino gauge [6].   This approach is completely
straightforward, and independent of the  details of the
particular superspace under consideration.  The observation 
of [2] is that standard methods in Riemannian geometry for
constructing  such coordinate systems [7] can be applied in
order to obtain component actions from superspace actions
much more efficiently than older methods. (However, this
method is not as efficient as one might hope, since it
requires evaluation  of the entire vielbein for the purpose of
determining just the measure --- the vielbein
superdeterminant.)  

On the other hand, the ``ectoplasmic" method of [3]
requires detailed knowledge of the properties of the
relevant superspace.  For example, for the standard case of
simple supersymmetry in four space-time dimensions,  the
existence and properties of chiral superspace must be
understood.   However, since the method works directly
with densities (as differential  forms), knowledge of  the
explicit measure is avoided.   Also, the strong dependence of
this method on the existence of ``subsuperspaces" like
chiral superspace makes it useful for studying them in the
less-known cases of extended supersymmetry.  (The use of
such spaces simplifies calculations  for the normal
coordinate method as well.)

In this letter we apply both methods to  the example of
four-dimensional N=1 supersymmetry, obtaining locally
supersymmetric  component  lagrangians for old minimal
supergravity.  (Similar results can be obtained for other
versions of supergravity.)  Its torsions and curvatures are
given by (we use the notation ${a} \equiv \a
\ad$ and in expressions such as $\psi_a{}^\a$ that we
encounter,
 summation over $\a$ is to be understood)
$$ {\ \ \ \ \ \ \ \,\ } [ \de_{\a} \ ,\  \de_{\b} \} \ =\  - 2
\,{\Bar R} {\cal M}_{\a 
\b} \ \ \ ,\ \ \  [ \de_{\a} \ ,\  {\Bar \de}_{\dot \b} \} \ =\  i  \,
\de_{\a
\dot 
\b}\ \ \ , {\ \ \ \ \ \ \ \ \ \ \ \ \ \ \ }{\ \ \ \ \ \ \ \ \ \ \ \ \ \ \
\ }$$
$$[ \de_{\a} \ ,\  \de_{b} \} \ =\  -\  i C_{\a \b} \,
[\  {\Bar R}
\, {\Bar
\de}_{\dot \b} \ -\  G^{\g} {}_{\dot \b} \, \de_{\g} \ ]  \ -\  i
(\, {\Bar
\de}_{\dot
\b} {\Bar R} \,) {\cal M}_{\a \b}{\ \ \ \ \ \ \ } {\ \ \ \ \ \ \ } $$
$$ {\ \ \ \ \ \ \ \ \ \ \ \ \ } +\  i  C_{\a \b} \, [\   {\Bar
W}_{\dot \b
\dot
\g} {}^{\dot \d}{\Bar {\cal M}}_{\dot \d }{}^{\dot \g} \ -\ 
(\de^{\g} G_{\d
\dot \b} ) {\cal M}_{\g }{}^{\d} \ ] \ \ \ , {\ \ \ \ \ \ \ \ \ \ \ \
}$$
$$[ \de_{a} \ ,\  \de_{b} \} \ =\  \{ \, [\ 
C_{\dot
\a
\dot \b} {W}_{\a \b} {}^{\g} \ +\   C_{\a \b} (\,  {\Bar
\de}_{\dot \a}  G^{\g} {}_{\dot
\b} \,)   \ -\  C_{\dot \a \dot \b} (\, \de_{\a} R  \,)\, \d_{\b}
{}^{\g} \ ] 
\de_{\g} {}\, \,\, \, \,$$
$${\ \ \ \ \ \ \ \ \ \ \ \ \ \ }\,\, \,  +\ i  C_{\a \b} G^{\g} {}_{
\dot
\b}
\de_{\g 
\dot \a} \ -\  [\, C_{\a \b} (\, {\Bar \de}_{\dot \a} \de^{\d} 
G_{\g 
\dot \b} \, )   {\ \ \ \ \ \ \ \ \ \ \ \ \ \ }$$ 
$${\ \ \ \ \ \ \ \ \ \ \ \ \ \ \ \ }\, \,\, \, \, \,  \ -\  C_{\dot \a
\dot
\b} (\,
\de_{\a}   {W}_{\b \g} {}^{\d} 
\ +\   (\, {\Bar \de}{}^2 {\Bar R} \ +\  R  {\Bar R} \,)  C_{\g \b} 
\d_{\a} {}^{\d} ) \ ] {\cal M}_{\d} {}^{\g} \, \} $$
$$ \ +\  {\rm h}. \ {\rm c}.  {\ \ \ \ \ \ \ \ \ \ \ \ \ \ \ \ \ \ \ \
\ \
\ \ \ \ \
\ \ \ \ \ \ \ \ \ \ \ } \,
\(1.1)$$
 We note for future reference that a covariantly chiral scalar
can be expressed in terms of a general scalar:
$$ {\Bar \de}_{\Dot \a}\Phi \ =\  0 \quad\Rightarrow\quad  
        \Phi \ =\   (\, {\Bar \de}{}^2 \ +\  R  \, )\, \Psi \, .  \(1.2)
$$
 Interestingly, the relevance of chiral lagrangians for
four-dimensional, N = 1 superspace follows from the
ectoplasmic approach, while the above relation  between
chiral and general scalars follows from the normal
coordinate  approach,  showing their complementarity. 

\sect{2. Fermionic Normal Coordinates}

Normal coordinates $y^A$ around an ``origin'' $z^M$ can be
defined by  Taylor expanding the metric or vielbein (and
other gauge fields) with respect to (some of) the
coordinates, such that the coefficients are field strengths
(and their derivatives).  This covariant expansion is
achieved by performing finite parallel transport from the
arbitrary point $z^M$.  An infinitesimal transport with
parameter $y^A$ is exponentiated to yield the finite one.
Taylor expansion of the exponential yields an explicit
algorithm for the desired coordinate expansion.  

In general, we expand (choose a normal gauge)  with
respect to a  subset of the coordinates:    We divide them 
into sets $(z^i, z^s)$ and $(y^i, y^s)$. After using the
algorithm we set $z^s=y^i =0$ (more generally the
$z^s$ can be set equal to some arbitrary constants) and use
$(z^i, y^s)$ as   new coordinates. The surviving $y$'s are the
normal-gauge-fixed coordinates, while the remaining $z$'s
are still arbitrary coordinates. 
  Such gauges are useful, e.g., for ``compactification",
where expansions are made in some of the coordinates,
while coordinate invariance is still desired in the remaining
coordinates. A familiar case is that of Riemann normal
coordinates, where we set all the $z$'s to vanish, and keep
all the $y$'s as our new coordinates.  A more relevant
example is Gaussian normal coordinates, where we choose
$y$ to be a single timelike coordinate, and $z$ the spacelike
coordinates, by setting $z^0=y^i=0$.  This construction then
gives the timelike gauge $g_{m0}=\eta_{m0}$, fixing the
time coordinate while leaving the space coordinates
arbitrary.  For  covariant component expansions in
supersymmetry, the idea is to fix  the fermionic  
coordinates, while maintaining coordinate invariance in the
bosonic coordinates.

Once the fermionic expansions have been obtained, they can
be applied to the superspace action.  Consider, for example,
the derivation of a covariant component action from a
superspace action  of the form
$$ S \ =\  \int d^4 x\ d^4 \theta\ E^{-1}L \(2.1) $$ 
 where $E^{-1} = sdet\ {E_M}^A$, and ${E_M}^A$ is the
vielbein.  We  first apply the algorithm to $E^{-1}L$ with
respect to
 $z^M = (z^m, z^\mu)$ and $y^A = (y^a, y^\alpha)$, and
then set $y^a=z^\mu=0$, 
 identifying $y^\a$ as the fermionic coordinates $\theta^\a$
in the Wess-Zumino gauge and  $z^m$ as the bosonic
coordinates $x^m$, still arbitrary with respect to spacetime
coordinate transformations.   (Generically we use
lower-case Greek letters for both dotted and undotted
spinor indices,
 wherever no distinction needs to be made.) 
  Integration over $\theta$ can then be performed as in flat
superspace, by picking out the highest-order terms in
$E^{-1}L$.  The expansion of $E^{-1}$ automatically gives the
usual factor of $e^{-1} = det\ {e_m}^a$.  As mentioned
earlier, $z^\mu$ can be set to an arbitrary constant  
(distinct from the integration variable
$\theta^\a=y^\a$); the result for the action is independent
of it.

However, superspace integrations are often  performed over
``subsuperspaces" parametrized by the usual spacetime
coordinates plus a subset of fermionic coordinates, for
example 
 (anti)chiral superspace for D=4, N=1.  (In fact, the use of
such subsuperspaces can be avoided only in degenerate
cases for D$<$4.  For massless theories, only one quarter of
the off-shell supersymmetries are physical, except when
there are fewer than four to start with.)  In such cases
normal coordinate expansions can be used for two purposes:
(1) reducing a  full superspace action, e.g. $S= \int d^4 x\
d^4 \theta\ E^{-1}L$
 to a subsuperspace action, e.g.  $ {\cal S}= 
\int d^4 x\ d^2 \theta\ {\cal{E}}^{-1}{ L_{ch}}$  , and (2)
deriving the  component expansion of the subsuperspace
action, e.g. ${\cal S} = \int d^4 x e^{-1}{\cal L}$. 

 In fact, these are the two steps we use  in practice to
evaluate the component expansion of  a full  superspace
action.  We  can interpret ${\cal S}$ as being obtained from 
$S$,  where we have
 expanded in (and integrated over) only with respect to the
$\bar{\theta}^{\Dot\alpha}$.  
  (Of course, in any situation coordinates can be integrated
out one at a time, but the result will not always be simple. 
Reduction to subsuperspaces produces a manifestly
covariant result only when scalars can be defined on such
spaces:  For example, chiral scalars exist in curved 4D, N=1
superspace.)  The procedure in both steps is the same; the
only difference is the choices of the various sets of
coordinates ($z$'s and $y$'s).  For our chiral superspace
example, which we will discuss in more detail below, the
first step parallel transports with respect to
$\bar\theta^{\Dot\alpha}$, dividing up the superspace
coordinates as
$(z^m,z^\mu;\bar y^{\Dot\alpha})$, while the second step
transports with respect to $\theta^\alpha$, dividing up the
coordinates as
$(z^m,\bar z^{\Dot\mu};y^\alpha)$ where we now explicitly
distinguish dotted and undotted indices. 

We first give the algorithm for finite parallel transport by
exponentiating an infinitesimal transformation.  We apply
repeatedly the rules for variation of the occurring
quantities (see [2] for details):
$$ \delta y^A = 0\, ,\quad   \delta {\cal T} = y\cdot\de {\cal
T}\, ,\qquad
$$
$$  \qquad \qquad    \delta E^A = Dy^A +y^C E^B T_{BC}{}^A\,
,\quad    
        \delta (Dy^A) = y^B y^C E^D R_{DCB}{}^A \, ,   \qquad    
\(2.2) $$
 where $\cal T$ is any tensor (including $T$ and $R$), which
is a function of just $z^M$, and
$$ E^A \equiv dz^M E_M{}^A (z) \quad ,\quad   
        Dy^A \equiv E^B \de_B y^A
        = dy^A -y^B E^C \omega_{CB}{}^A(z) \, .  \(2.3) $$
 The transformation of the tensor identifies $y^A$ as the
translation parameter, while $\delta y^A=0$ is the geodesic
condition.  (The other transformations follow from that of
the tensor, since then
$\delta\de=[y\cdot\de,\de]$.)

Applying the above algebra mechanically, we evaluate the
transformed vielbein, breaking it up into a part proportional
to  $E^B(z)$ and a part proportional to $Dy^B$:
$$ E'^A (z;y) \equiv \sum_{n=0}^\infty \f1{n!}\delta^n E^A 
        = E^B(z) F_B{}^A +(Dy^B)G_B{}^A \, ,\(2.4) $$ 
 where $F$ and $G$ depend explicitly only on tensors ${\cal
T}(z)$ ($T$ and $R$ and their covariant derivatives), and on
$y^A$.  For  the present purpose  we  need only the terms of
order $n=0, 1$ (given above), 2:
$$ \d^2E^A = y^B y^C E^D R_{DCB}{}^A 
       +y^C(Dy^B +y^E E^D T_{DE}{}^B)T_{BC}{}^A
        +y^C E^B y^D \Del_D T_{BC}{}^A \, . \(2.5) $$ and the
part of
$\d^3E^A$ proportional to $Dy^B$.

The first step in applying the algorithm is to divide up the
$z^M$ and $y^A$ coordinates into complementary sets.  The
same procedure applies for expansion about any subset of
the coordinates; only the ranges of the indices change.  To
simplify notation, we will use indices corresponding to the
special case of expansion of the full superspace over all the
fermionic coordinates, with the understanding that
appropriate modifications can be made for other cases.  We
thus evaluate at
$$ z^\mu \ =\ 0 \ ,\  y^a \ =\  0 \, ,  \(2.6)$$
 (and similarly for $dz^\mu$ and $dy^a$):  This is the
expansion in the fermionic normal coordinates $y^\alpha$
about the bosonic hypersurface coordinatized by $z^m$. 
The left-hand-side of  (2.4) then gives the complete vielbein
in all superspace
$$ E^A (z^m,y^\alpha) \ =\  E'^A (z^m,0;0,y^\alpha) \, ,\(2.7)
$$
 while on the right-hand-side we have only $y^\alpha$ (and
$dy^\alpha$),
${\cal T}(z^m,0)$ (and $\omega(z^m,0)$, but the connection
cancels
 out in any Lorentz invariant quantity and need not be
considered  in the expansion of the action), and only the
nontrivial part of the vielbein (since $dz^\mu=0$):
$$ E^A(z^m,0) \ =\  dz^n E_n{}^A(z^m,0) 
        \ =\  E^b(z^m,0)\check E_b{}^A(z^m,0), $$
$$ \check E_a{}^B(z^m,0) = (\delta_a^b, -\psi_a{}^\beta )
\(2.8)$$
 (In the first equation we have factored out  the component
vielbein;  the second equation is consistent with the
standard definition of the gravitino. We do not need to
explicitly   fix  a  gauge and use
 $E_\m{}^A$.)  We thus have
$$ E^A (z,y) \ =\  E^b(z,0)\hat E_b{}^A(z,y)
        +(Dy^\beta)\hat E_\beta{}^A(z,y); $$
$$ \hat E_b{}^A = \check E_b{}^C(z,0) F_C{}^A(z,y),\quad
        \hat E_\beta{}^A = G_\beta{}^A(z,y) \(2.9) $$
 From the normal coordinate expansion, we read off the
vielbein components, and evaluate the superdeterminants
$$ E^{-1}(z,y)  \equiv sdet\ E_M{}^A(z,y)\ =\  \tilde E^{-1}
\hat E^{-1}; $$
$$   \tilde E^{-1} \equiv sdet\ E_m{}^a(z,0),\ 
        \hat E^{-1} \equiv sdet\ \hat E_A{}^B(z,y) \(2.10) $$

In ref. [2] we have given a number of examples of the
procedure. Here we summarize the case of four-dimensional
N=1 minimal supergravity. We consider evaluation of a 4D
N=1 superspace integral over
$d^4\theta = d^2\theta d^2\bar\theta$.  This would require
expansion up to fourth (and  partly fifth) order  in all the
fermionic coordinates and therefore
 we consider instead the two-step procedure described
above. In the first step we do only the
$\bar{\theta}$ integration, by expanding with respect to
$\bar{y}^\ad$ and evaluating at
$$  \bar{z}^{\dot{\mu}} =0 \ \ \ \ ,\ \ \  (y^a,y^\a)=0  
\(2.11) $$
 (where we now explicitly distinguish between dotted and
undotted spinors).  We now need to expand $E^A$ only to
second (and partly third) order.  We find
$$ \hat E_B{}^A = \pmatrix{
        \d_b^a -i \bar y^\ad \psi_b^\a -\f12 \d_b^a \bar y^2 R
                & -\psi_b^\a +i \bar y^\gd C_{\gd\bd}  \d_\b^\a
R+\cdots
                & - \psi_b^\ad +\cdots \cr
        -i\bar y^\ad \d_\b^\a  +\cdots
                & \d_\b^\a +\bar y^2 \d_\b^\a R
                & \cdots \cr
        0 & 0 & \d_\bd^\ad + \f12\d_\bd^\ad \bar y^2 R \cr}
\(2.12) $$
 We have written  only the relevant terms.  Evaluating the
superdeterminant we find
$$ E^{-1} \ =\  \tilde E^{-1}\, (\, 1 \ -\  \bar y^2 R \, )   
\(2.13)$$
 where $\bar y^2 = \f12 \bar{y}^{\Dot\alpha}
\bar{y}_{\Dot\alpha}$ and
$$ \tilde E^{-1} \ =\  sdet \pmatrix{ E_m{}^a & E_m{}^\alpha
\cr
        E_\mu{}^a & E_\mu{}^\alpha \cr}  \(2.14) $$
 is the measure of the chiral subspace.  (Explicit evaluation
by the usual methods [1] shows it equals the usual measure
$\phi^3$ in terms of the chiral compensator [4]).  We also
expand the scalar lagrangian
$$ L(\bar y) \ =\  (\, 1 \ +\  \bar
y^{\Dot\alpha}\de_{\Dot\alpha} 
        \ -\  \bar y{}^2\, \Bar\de{}^2 \,)\, L(0)     \(2.15)$$
Performing the
$\bar y$ integration, we find
$$ S \ =\  \int d^4 x\ d^4\theta\ E^{-1}L \ =\  \int d^4 x\
d^2\theta\ 
\tilde E^{-1}(\Bar\de{}^2 \, + \, R)\, L    \(2.16)$$
 The normal coordinate method thus gives the solution to
the chirality condition as well as the chiral integration
measure.  Note that the
$\psi$ terms cancelled:  This expresses the covariance of
the chiral subsuperspace (i.e., of chiral scalar superfields).

The same procedure can be applied to the next step.  The
coordinate restriction is now the antichiral one
$$ z^\mu \ =\ 0 \ ,\ (y^a, \bar y^{\Dot\alpha}) \ =\  0   
\(2.17)
$$
 We first need to evaluate $\tilde E$ of (2.14), now treated
as a function of the new $z^M$ and $y^A$.  We can use the
result of the previous calculation by: (1) replacing (2.12)
with the hermitian conjugate (effectively just switching
dotted and undotted indices), and (2) deleting the second
row and column before taking the superdeterminant, so we
get the contribution to the smaller superdeterminant of
(2.14). The result is
$$ \tilde E^{-1} \ =\  e^{-1}\, [ \, 1 \, -\, iy^\a \bar\psi_{a}{}^\ad 
\,-\, y^2 \, ( \, 3 \,\bar R \,+\, \f12 C^{\a \b}\, \bar\psi_{a} {}^{(\ad }
\bar\psi{}_{b}{}^{\bd )}
 \,) \ ]   \(2.18) $$ where $e^{-1}=det\ e_m{}^a$ is the new
$\tilde E^{-1}$ part of this subdeterminant, and the rest is
the new $\hat E^{-1}$.  The expansion of the lagrangian is
similar to the previous case (just switching
$\bar y\to y$), giving the final result
$$ S \ =\  \int d^4 x\ d^4 \theta\ E^{-1}L
        \ =\  \int d^4 x\ {\rm e}^{-1}{\cal D}^2 ({\Bar \de}{}^2
\,+\, R ) L
        \ =\   \int d^4 x\ {\rm e}^{-1}{\cal D}^4 L  \(2.19) $$
 where we have defined a superdifferential operator, the
``chiral density projector" ${\cal D}^2$, which (for the
present case of  old minimal supergravity) takes the form
$$ {\cal D}^2 \ \equiv \  \nabla^2 \,+\, i  {\bar \psi}{}^{a}{}_{\Dot
\a} 
\nabla_{\a} \, +\, 3 \,{\bar R} \, +\, \f12 C^{\a \b} {\bar
\psi}_{a} {}^{( \Dot \a} \, {\bar \psi}{}_{b}{}^{\Dot \b
)}     \(2.20) $$
 and the general density projector 
${\cal D}^4\equiv{\cal D}^2(\Bar\de {}^2+R)$.  We could use 
instead the complex conjugate $\bar{\cal D}^4=\bar{\cal
D}^2(\de^2+\bar R)$, which differs only by a total
space-time derivative.

In the calculation above we have set the fermionic
coordinates $z^\m $ to zero, but they could be fixed at any
other value since  the action is  independent of the
fermionic coordinates. Independence from
$\theta^\m$ is equivalent to supersymmetry invariance; in
superspace, supersymmetry transformations are formulated
as coordinate transformations.  This fact is the basis of the
``ectoplasmic" method of superspace integration, which we
will now describe.

\sect{3. Ectoplasmic Subintegration}

Our goal is to use superfield methods to construct {\it
locally}  supersymmetric {\it component} actions. 
Component actions are written  as integrals over
space-time, while superfield actions are written as 
integrals over superspace.  Since space-time is a subspace
of superspace,  it is natural to consider the same approach
that is used for integration  over subspaces of spaces with
only commuting coordinates.  For example,  when
considering integrals over three-dimensional hypersurfaces
in  four-dimensional space-time, we integrate the
component of a vector normal  to the surface.  The condition
that the integral be independent of the  choice of
hypersurface is the constraint that the vector be a
conserved  current. In the same fashion if  we consider an 
integral over a  space-time hypersurface in superspace,
local supersymmetry invariance translates into
independence of the integral from the $\theta$ variables in
the integrand and imposes similar  ``conservation''
conditions on the  latter.  Solving these conditions allows us
to express the integrand in  a manifestly  locally
supersymmetric component form.

This analysis can be applied to spaces without a metric.  In
the case under discussion it is necessary to integrate a 
differential three-form, or covariant third-rank
antisymmetric tensor, and  the hypersurface element itself
is described by a contravariant third-rank  antisymmetric
tensor.  A conserved ``charge'' in this case takes the form
$$ Q \ =\  \int dx^{m} \wedge dx^{n} \wedge dx^{p} \  J_{p n
m}  \ \  . \(3.1) $$

Time independence of the integral implies the conservation
law;
 the 3-form current must be closed (curl-free) but  not
exact (i.e. not globally the curl of a 2-form):
$$  {dQ\over dt} = 0 \quad\Rightarrow\quad
        \partial_{[m}J_{n p q]} \ =\  0\, ,\quad 
        \delta J_{m n p} \ =\  \partial_{[ m} \lambda_{ n p]} \ \
\  .  \(3.2) $$
 (In differential form notation, we write simply $Q=\int J$
satisfying
$dJ=0$, $\delta J=d\lambda$.)  We assume that $J$ and
$\lambda$ vanish  at the boundaries.  

We can also interpret the conservation law as saying  that
$J$ is a gauge field (with gauge parameter $\lambda$)
whose field strength vanishes.  However, this is not a gauge
field in the usual sense of de Rham cohomology (cohomology
of $d$):  We consider only local functionals of the fields,  not
arbitrary functions of the coordinates, and the gauge
invariance is  just the statement that we drop total
derivatives in the integrand.  We can evaluate the charge
$Q$ at $t=0$ for convenience, but the conservation law was
derived to make that unnecessary.

This analysis can be applied without modification to
component actions, interpreted as integrals  of superforms
over a space-time ``hypersurface"
 in superspace.  We thus evaluate the integral in $D$ 
space-time dimensions:
$$  S \ =\  \int dx^{{m}_1} \wedge ... \wedge dx^{{m}_D} \  
J_{{m}_D...{m}_1} \ =\  \int d^D x \  \f1{D!}\epsilon^{{
m}_1...{m}_D} J_{{m}_D...{m}_1} \ \ \  .
\(3.3) $$ The integrand, $J_{{m}_D...{m}_1}$, is in general a
function of superfields depending on both space-time and
fermionic variables. The ``conservation law" is now 
$$ \partial_{[M_1}J_{...M_{D+1})} \ =\  0\, , \quad 
\delta J_{M_1...M_D} = \partial_{[M_1} \lambda_{...M_D)}\,
.      \(3.4)
$$ Let us emphasize that the indices that appear in these
equations are ``holonomic'' or ``curved'' super-vector
indices.  Separating out the  bosonic and fermionic parts of
the indices we have
$$ \partial_{[{m}_1}J_{...{m}_{D+1}]} \ =\ 
0,\quad\partial_{\mu} J_{{m}_1...{m}_D} 
-\f1{(D-1)!}\partial_{[{m}_1|}J_{\mu|{ m}_2...{m}_D]}\ =\ 
0,...\,.  \(3.5) $$
 The first equation is trivial, since there are more than $D$
$D$-valued space-time indices antisymmetrized.  (Thus,
there are no conditions for the nonsupersymmetric case.) 
After using the gauge invariance to pick a convenient gauge,
the remaining equations  in (3.5) can be solved to give
$$ \partial_{\mu}J_{{m}_1...{m}_D} \ =\  0 \(3.6) $$
 Thus, {\it in this particular gauge} the integrand is in fact
independent of the fermionic variables and can then  be
expressed as an integral (or derivative) of some object $L$
with respect  to all the anticommuting coordinates. 
$$ J_{{m}_1...{m}_D} =
        \int d^{N_F} \theta\ L_{{m}_1...{m}_D}  \ \ \  . \(3.7) $$
where $N_F$ is the number of fermionic coordinates in the
superspace. In writing the solution in (3.7) we recall that
the Berezinian definition of the Grassmann integral implies
that it is equivalent to differentiating with respect to all of
the fermionic coordinates.    The original integral then is 
converted into the standard integral over all superspace of
{\it some} superspace lagrangian.

Although this  result may be sufficient for flat (super)space,
it does not result  in a covariant expression in the presence
of supergravity. (In what follows we return to a general
gauge so that (3.6) is  necessarily  not satisfied.) The
simplest way to solve the conservation law covariantly is to
convert from curved to flat indices, since the superspace of
supergravity needs the tangent space for its definition.  The
conversion is
$$  J_{{m}_1...{m}_D} \ =\ E_{{m}_D}{}^{A_D} ...\, \,  E_{{
m}_1}{}^{A_1}J_{A_1...A_D} \ \ \  .  \(3.8) $$
 and similarly for $\lambda$.  Note that, as for the normal
coordinate method, the only parts of the vielbein that
appear are  the nontrivial parts, namely $ e_{m}{}^{a}(x,
\theta) \equiv E_{m}{}^{a}$ and $\psi_{m}{}^\alpha (x,
\theta )\equiv -E_{m}{}^\alpha$ that contain  at  $\theta =0$
the graviton, and the gravitino, respectively.   (This is the
same expansion as in (2.8).)  We emphasize that  the result
of the integration is  $\theta$-independent by construction
so that  the specific choice  of a particular hypersurface,
e.g.  at $\theta=0$, is never needed explicitly  although it
may be convenient.

We now need to solve the conservation constraints.   We
convert the indices of (3.4)  by appropriately multiplying by
the vielbein $E_A {}^{M}$ so that
$$ \f1{D!}\nabla_{[A_1}J_{A_2...A_{D+1})} \, - \,
\f1{2(D-1)!}\,  T_{[A_1 A_2|}{}^B J_{B|A_3...A_{D+1}]}  \ =\   0
\ \ \  ,  \(3.9) $$
$$  \delta J_{A_1...A_D} \ =\ 
\f1{(D-1)!}\nabla_{[A_1}\lambda_{A_2...A_D)}
        -\f1{2(D-2)!}\, T_{[A_1 A_2|}{}^B
\lambda_{B|A_3...A_D]}\ \ \  . 
\(3.10) $$
 Using the explicit form of the torsion in terms of the
irreducible tensors of supergravity, and especially
$T_{\alpha\beta}{}^c\sim 
\gamma_{\alpha\beta}^c$, the pieces of $J$ with more
vector indices  can be recursively expressed in terms of
those with more spinor indices.   The form of the solution
depends on the particulars of the supergravity  under
consideration.  The general result is that a certain part of
$J$  is a scalar, which may satisfy some (spinorial)
differential constraint;  parts of $J$ with more spinor
indices vanish, those with fewer are  expressed as spinorial
derivatives of that scalar (with also torsion  terms).  The
scalar is the superspace lagrangian, and its constraints (if 
any) define the kind of superspace.  

To illustrate how all of this comes together, we examine the
case of  local $N=1$ supersymmetry in the case of $D=4$. 
The expanded form  of (3.8) is given by
$$ J_{{m}_1 ... {m}_4} \ =\   e_{{m}_1} {}^{f}\,  \cdots \,
e_{{m}_4} {}^{k} \  \Big[ 
\  J_{f \, g \, h \, k}  \ -\  (\, 
\f 1{3!} \psi_{[ f |}{}^{\a} J_{\a \, | g \, h \,  k] } \ +\  \f 14
\psi_{[ f  |}{}^{\a} \psi_{| g |} {}^{\b}  J_{\a \, \b \, | h \, k] } 
 \ +\  {\rm {h.\, c.}} \ )  ~ $$ 
$$ \ \ \ \ \ \ \ \ \ \ \ \ \ \ \ \ \ \ \ \ \ \ \ \ \ \ \ \ \ \ \ \ \ \ \
\ -~ \f 1{2} \psi_{[ f |}{}^{\a} {\Bar \psi}_{ | g |}{}^{
\dot \b} J_{\a \, {\dot \b} \, | h \, k] } ~ +\  (\, 
\f 1{3!} \psi_{[ f |}{}^{\a} \psi_{| g |}{}^{\b} \psi_{|  h
|}{}^{\g}  J_{\a \,\b \,  \g \, | k] }  \ +\  {\rm  {h.\, c.}} \ ) $$
$$ \ \ \ \ \  \ \ +\   \f 12 (\, \psi_{[ f |}{}^{\a} 
\psi_{| g |}{}^{\b} {\Bar \psi}_{| h |}{}^{\dot \g}  J_{\a \, 
\b \,  {\dot \g} \, | k] }  \ +\  {\rm {h. \, c.}} \ ) $$
$$ \ \ \,\ \ +\  (\, \psi_{ f }{}^{\a} 
\psi_{ g }{}^{\b} \psi_{h} {}^{\g} \psi_{k}{}^{\d} J_{\a 
\, \b \, \g \, \d }   \ +\  {\rm {h.\, c.}} \ ) $$
$$ \ \ \ \ \ \,\ \ \ \ \ \ +\  (\, \f 1{3!} \psi_{[  f |}{}^{\a}
\psi_{| g |}{}^{\b} {\psi}_{| h |}{}^{\g} {\Bar 
\psi}_{| k ]}{}^{\dot \d}  J_{\a \, \b \, {\g} \, {\dot \d} }  
 \ +\  {\rm {h.\, c.}} \ )  $$
$$  \,\ \ +\   \f 14 \psi_{[ f |} {}^{\a} \psi_{| g |}{}^{\b} {\Bar
\psi}_{| h |}{}^{\dot \g} {\Bar \psi}_{| k ]} {}^{\dot \d}  J_{\a
\, \b \,  {\dot \g} \, {\dot \d} }  \ \Big] \ \ \ .
\(3.11) $$

 In writing this we note the following: on the
right-hand-side of the equation appear the components of
the vielbein that we have denoted above as $e_{m}{}^{f}(x,
\theta)$  and $-\psi_{f} {}^{\a}(x,
\theta)$.  However, since we know that $ \int J_{{m}_1 ... 
{m}_4}$ is $\q$-independent,  we can evaluate it at $\q
= 0$ and hence  use the values of the vielbein and
$J_{A_1...A_4}$ at $\q = 0$ in terms of  the ordinary
component vielbein and gravitino.

  The solution to the   constraints is given simply by the
supercovariantization of the rigid results [8] expressing all
the nonvanishing components of the superform in terms of a
covariantly chiral scalar superfield ${\cal J}$. Up to total
derivative terms that give no contributions to the action
integral
$$   J_{\a \b \g D} \ =\   J_{\ad\b\g D} \ =\   J_{\ad\b  c d} \
=\   0\, ,\quad J_{\a\b c d} \ =\   iC_{\Dot \g \Dot \d} C_{\a
(\g} C_{\d ) \b} {\bar {\cal J}}\, ,\quad
{\Bar\nabla}_{\ad}{\cal J} \ =\   0\ \
\,   $$
$$ J_{\a b c d} =  -i \e_{a b c d}
{\Bar\nabla}{}^{\ad}{\bar{\cal J}}\ \ \  ,\quad J_{\Dot\a b c 
d} \ =\  -i\e_{a b c d}{\nabla}^{\a}{\cal J}\ \ \  , 
$$
$$  J_{a b c d} \ =\   \e_{a b c d}\left[ 
\ ( {\nabla}^2 \,+\,        3{\Bar R}){\cal J} \ +\ 
({\Bar\nabla}{}^2 
\,+\, 3R){\bar{\cal J}}\  \right] \ \ \  ,  \(3.12) $$
 where
$$ \e_{abcd} \ \equiv\   i(C_{\a\d}C_{\b\g}C_{\Dot\a\Dot\b}
        C_{\Dot\g\Dot\d} \,-\, C_{\a\b}
        C_{\g\d}C_{\Dot\a\Dot\d}C_{\Dot\b\Dot\g})\ \ \  . 
\(3.13) $$

Substituting (3.11) and (3.12) into (3.3) yields the result
$$ S \ =\  \int d^4 x\  {\rm e}^{-1}{\cal D}^2{\cal J}\ +\  {\rm
h.c.}
\ \ \ , \(3.14) $$
 with
$$ {\cal D}^2 \ \equiv \  \nabla^2 \,+\, i  {\bar
\psi}{}^{a}{}_{\Dot
\a} 
\nabla_{\a} \, +\, 3 \,{\bar R} \, +\, \f12 C^{\a \b} {\bar
\psi}_{a} {}^{( \Dot \a} \, {\bar \psi}{}_{b}{}^{\Dot \b )}    
\(3.15) $$
 as in (2.20). Being $\q$-independent, the right hand side
may be evaluated at $\q = \bar\q = 0$  without  loss of
generality. This result agrees with (2.19); solving the 
chirality condition as in (1.1), we find
$L=\Psi +\bar\Psi$.   Even  without (2.19), it is clear we can
identify (3.14) with the result of integration over $\theta$,
since the chiral projector is the covariantization of the
flat-space integration.  (If there were an ambiguity
associated with ``nonminimal coupling", it would have
shown up as an ambiguity in solving the constraints.) 

In conclusion,  we  emphasize that the ``ectoplasmic''
method for the construction of local density projectors is a
genuinely new approach to the problem of integrating over
the supermanifolds appropriate for describing locally 
supersymmetric theories. At no stage did we require any
knowledge of the superspace measure, e.g. $( {\rm
sdet}E)^{-1}$.  The superspace lagrangian and the form of
the component  action is determined solely from the
existence of superforms and the solution of the constraints
they satisfy.

\refs

\Item 1 S.J. Gates, M.T. Grisaru, M. Ro\v{c}ek and W. Siegel,
{\it Superspace} (Addison-Wesley, 1983).

\Item 2 M.T. Grisaru, M.E. Knutt-Wehlau and W. Siegel, {\it A
Superspace Normal Coordinate Derivation of the Density
Formula}, Brandeis Univ. Preprint BRX-TH-422 (also
ITP-SB-97-56 and  WATPHYS-TH-97-10), hep-th/9711120.

\Item 3 S.J. Gates, Jr., {\it Ectoplasm Has No Topology: The
Prelude}, Univ. of MD preprint UMDEPP 98-13,(Sept. 1997)
hep-th/9709104.

\Item 4 I.N. McArthur, { Class. Quantum Grav.} {\bf 1} (1984)
233, {\bf 1} (1984) 245.

\Item 5 J.Atick and A. Dhar, { Nucl. Phys.} {\bf B284} (1987)
131.

\Item 6 J. Wess and B. Zumino, { Nucl. Phys.} {\bf B78} (1974)
1.

\Item 7 M. Spivak {\it A comprehensive introduction to 
differential geometry} vol. II (Berkeley: Publish or Perish
Inc., 1979).

\Item 8 S.J. Gates, Jr., { Nucl. Phys.} {\bf B184} (1981) 381.

\end{document}